\documentclass[%
 reprint,
 bibnotes,
 amsmath,amssymb,
 aps,
floatfix,
]{revtex4-1}

\usepackage{graphicx}% Include figure files
\usepackage{dcolumn}% Align table columns on decimal point
\usepackage{bm}% bold math
\usepackage{float}
\usepackage{todonotes}
\begin{document}

\preprint{APS/123-QED}

\title{Base-pair mismatch can destabilize small DNA loops through cooperative kinking}% Force line breaks with \\
%new title needed

%Cooperative kinking in DNA cyclization
%Easy come, easy go: loops 

\author{Jiyoun Jeong}
\author{Harold D. Kim}%
 \email{Corresponding author.\\Email: harold.kim@physics.gatech.edu}
\affiliation{%
 School of Physics, Georgia Institute of Technology, 837 State Street, Atlanta, GA 30332-0430, USA
}%

\date{\today}% It is always \today, today,

\begin{abstract}
%Small DNA loops ($\sim$150 bp) are closely involved in crucial biological processes such as gene regulation and DNA packaging. Flexible defects in the double helix, such as mismatched base pairs, increase the rate of loop capture by allowing large-amplitude bending fluctuations around them. From the other direction, they are expected to decrease the rate of loop release by relieving the mechanical stress, but this idea has not been directly tested. To address how base pair mismatch affects the rate of loop release, we cyclized $\sim$100-bp DNA with sticky ends and measured the rate of decyclization with and without base pair mismatch using single-molecule Fluorescence Resonance Energy Transfer (FRET). Surprisingly, we found that the mismatch causes the decyclization rate to increase. This unexpected effect was contingent upon stackability between the sticky ends and was most prominent when the mismatch was at the midpoint of the DNA, furthest from the sticky ends. These findings indicate that base pair mismatch destabilizes the loop by increasing the local bending stress at the antipodal position of the loop. Our study thus represents a direct demonstration of an allosteric mechanism known as cooperative kinking and its impact on the kinetic stability of a macromolecular structure.  

Base pair mismatch can relieve mechanical stress in highly strained DNA molecules, but how it affects their kinetic stability is not known. Using single-molecule Fluorescence Resonance Energy Transfer (FRET), we measured the lifetimes of tightly bent DNA loops with and without base pair mismatch. Surprisingly, for loops captured by stackable sticky ends, the mismatch decreased the loop lifetime despite reducing the overall bending stress, and the decrease was largest when the mismatch was placed at the DNA midpoint. These findings show that base pair mismatch transfers bending stress to the opposite side of the loop through an allosteric mechanism known as cooperative kinking. Based on this mechanism, we present a three-state model that explains the apparent dichotomy between thermodynamic and kinetic stability of DNA loops. 

%\begin{description}
%use the optional argument of the \verb+\item+ command to give the category of %each item. 
%\end{description}
\end{abstract}

\pacs{Valid PACS appear here}% PACS, the Physics and Astronomy
\maketitle

%\tableofcontents

\begin{figure}[!b]
\begin{center}
\includegraphics[scale = 1]{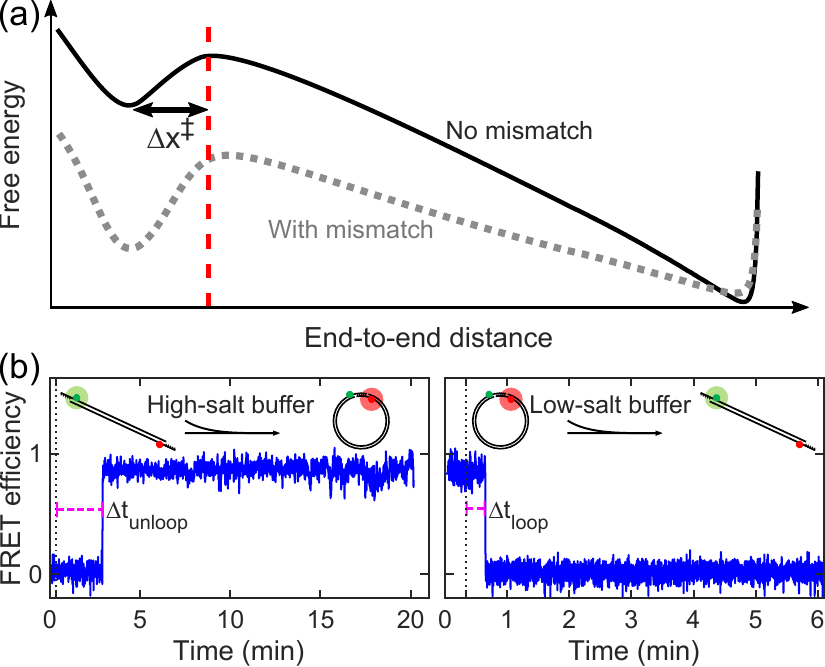}
\end{center}
\caption{\textbf{(a)} One-dimensional free energy landscape for DNA loop capture and release. The two minimum free energy states correspond to the looped and unlooped states. The transition state (vertical line) is separated from the looped state by a small distance $\Delta \textrm{x}^{\ddagger}$, which is equal to the capture radius. The base pair mismatch is expected to increasingly untilt the solid curve toward shorter end-to-end distances, which results in the dotted curve. \textbf{(b)} Typical FRET trajectories of a DNA molecule undergoing loop capture (left) and loop release (right). The DNA molecule labeled with Cy3 (green) and Cy5 (red) is in the low FRET state when unlooped, and in the high FRET state when looped. A sudden increase or decrease in NaCl concentration at the 20-second time point (marked by a vertical dotted line) triggers the transition.
}
\label{fig1}
\end{figure}

Cellular DNA is constantly exposed to the possibility of mispairing (i.e. non-complementary base pairing) \cite{Chatterjee2017}. Most commonly, mismatched base pairs result from base misincorporation during gene replication \cite{Kunkel2015} and heteroduplex formation between slightly different DNA sequences during homologous recombination \cite{Tham2016}. They can also arise from exposure to DNA damaging agents that modify nucleobases \cite{Cadet2013,Granzhan2014}. Due to less favorable base pairing and stacking \cite{SantaLucia2004}, mismatched base pairs can increase local flexibility of double-stranded DNA \cite{Rossetti2015, Sharma2013, Chakraborty2017}, and consequently the capture rate of tightly bent loops \cite{Dittmore2017}. For example, 1 to 3 bp-mismatch near the center of a short DNA fragment ($<$150 bp) was shown to increase the rate of DNA loop formation by one to two orders of magnitude \cite{Kahn1994,Vafabakhsh2012}. The kinetics of loop formation or capture is intuitively understood by a one-dimensional free energy curve with the end-to-end distance as a single reaction coordinate (Figure \ref{fig1}(a)). Base pair mismatch would reduce the mechanical work required to bring two distant DNA sites to proximity, more so for a shorter end-to-end distance. Therefore, the base pair mismatch would lower the transition state relative to the unlooped state (dotted line, Figure \ref{fig1}(a)). %Therefore, the effect of base pair mismatch can be interpreted as lowering the transition state relative to the unlooped state in the one-dimensional free energy diagram (Figure \ref{fig1}(a)).

Base pair mismatch is also expected to affect the breakage or release rate of small DNA loops that are captured by protein complexes \cite{Tardin2017} or by sticky ends of the DNA itself \cite{jeong2016single}. Looped DNA segments on the order of one persistence length are subject to a high level of mechanical stress; therefore, the free energy of the looped state is significantly lowered in the presence of the mismatch. According to the free energy diagram in Figure \ref{fig1}(a), the transition state, being at a slightly longer end-to-end distance by $\Delta \textrm{x}^{\ddagger}$, would be lowered to a lesser degree (Figure \ref{fig1}(a)). Therefore, the one-dimensional model predicts that the rate of loop release would decrease in the presence of base pair mismatch.

\begin{figure*}[!t]
\begin{center}
\includegraphics[scale = 1]{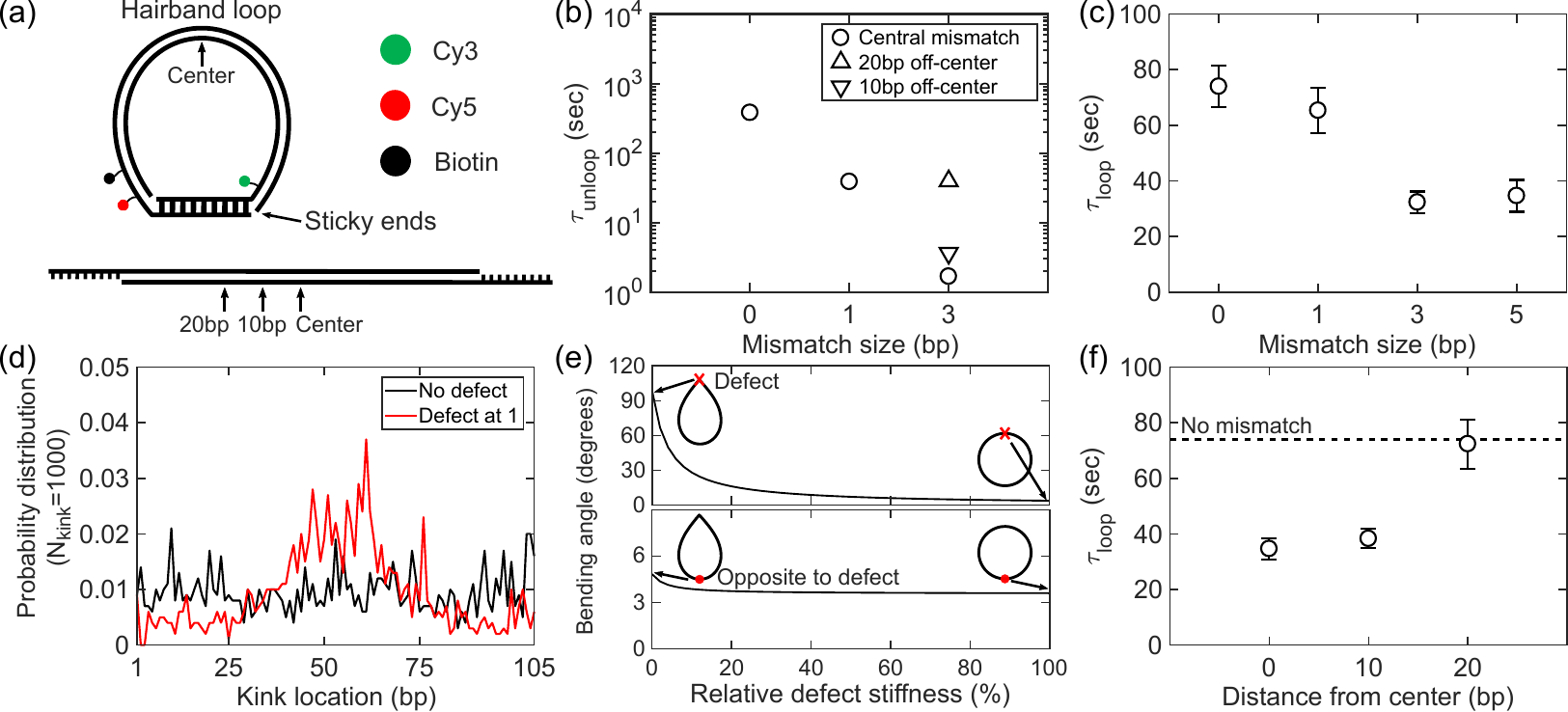}
\end{center}
\caption{\textbf{(a)} Schematic of a hairband loop captured by sticky ends. The schematic on top shows base-paired overhangs, Cy3 (green circle), Cy5 (red circle), and the biotin linker (black circle). In this geometry, the overhangs on opposite strands form a duplex that can stack at both nicks of the loop. Different positions of base pair mismatch tested in our experiments are marked on the linear form at the bottom. Only the bases on the overhangs are shown. \textbf{(b)} Loop capture time of the hairband molecules (108 bp) as a function of the central mismatch size (circles). Data with an off-center 3-bp-mismatch are also shown as triangles. The upright and flipped triangles represent the loop capture times for base pair mismatches placed at 20 and 10 bp away from the center of the molecule, respectively. Error bars, the standard errors of the mean, are smaller than the size of the symbols. \textbf{(c)} Hairband loop lifetime (loop release time) as a function of the central mismatch size. Error bars represent the standard errors of the mean. \textbf{(d)} Probability density of spontaneous kink positions along the coarse-grained minicircle (105 bp) with (red) and without a pre-existing flexible defect (black), which is placed at position 1. \textbf{(e)} Bending angle calculated from the minimum-energy conformation of a DNA minicircle (105 bp) with a defect. Top and bottom figures show bending angles at the defect and the site opposite to the defect, respectively, as a function of the defect stiffness relative to an intact base pair. The minimum-energy conformations of the two extreme cases of the defect stiffness (0 and 100\%) are also shown along the curves with the defect position marked by X. \textbf{(f)} Hairband loop lifetime as a function of the mismatch position (3-bp in size). For comparison, the horizontal dotted line shows the loop lifetime without the mismatch. Error bars represent the standard errors of the mean.}
\label{fig2}
\end{figure*}

Such prediction of mismatch-dependence seems plausible considering the success of the model in predicting the length dependence of loop capture and release rates. In the length regime where the free energy of loop formation is dominated by bending energy, increasing DNA length effectively reduces the tilt in the free energy curve because states at shorter end-to-end distances receive more stress relief, similar to the dotted line in Figure \ref{fig1}(a). This change predicts that loop capture and release rates measured at different DNA lengths would be anti-correlated; loops associated with higher mechanical stress are captured more slowly and released more quickly. This prediction has been confirmed for both DNA loops captured by Lac repressor \cite{chen2014modulation} and DNA loops captured by sticky ends \cite{le2013measuring,Le2014}. While increasing DNA length evens out the bending stress over the entire DNA molecule, the base pair mismatch tends to localize sharp bending. Therefore, the effect of base pair mismatch might be quite different from that of increasing DNA length. 

In this Letter, we investigated how base pair mismatch affects the stability of small DNA loops. As a model system for DNA loop capture and release, we used short double-stranded DNA molecules with sticky ends. To monitor loop capture and loop release events, we used the single-molecule FRET assay as previously published \cite{jeong2016single}. Briefly, DNA molecules labeled with Cy3 and Cy5 near their sticky ends were immobilized to a NeutrAvidin-coated glass surface through a biotin linker, and loop capture or release was triggered by exchange of buffers with different NaCl concentrations (see Supplemental Material for more details). 
%We used the single-molecule FRET assay \cite{Le2014,Jeo} to monitor DNA molecules undergoing cyclization or decyclization triggered by buffer-exchange. We prepared DNA molecules with sticky ends and labeled them with Cy3 (donor) and Cy5 (acceptor) fluorophores near the 5$^\prime$ end such that the looped and unlooped configurations were distinguished by the FRET efficiency (Figure \ref{fig1}(b)). These molecules were tethered on a temperature controlled flow-cell (20 $^\circ$C) that enabled exchange of buffers with different NaCl concentrations by which the stability of duplexed sticky ends was manipulated. 
The first transition times in the FRET signals ($\Delta t$) of $\sim$150 individual DNA molecules were collected. The mean of $\Delta t$ spent in the unlooped state before looping is defined as the loop capture time ($\tau_\textrm{unloop}$), and the mean of $\Delta t$ spent in the looped state before unlooping is defined as the loop release time or loop lifetime ($\tau_\textrm{loop}$). All DNA molecules used in this study were shorter than
150-bp, the length regime where the free energy of loop
formation is dominated by bending energy.

%to determine lifetimes of either the looped or unlooped state. In this experimental setup, sticky ends (9 nt each) were placed on the 5$^\prime$ end of the strands so that they could anneal in trans and stack upon each other (Figure \ref{fig2}(a)). In a previous study, we showed that this end-stacking, or equivalently nick closing, substantially increased the lifetime of the looped state. Based on the shape of the initially captured loop conformation, we term this loop as the hairband loop. The size of the hairband loop (including the length of annealed sticky ends) was carefully chosen to allow nick closing without excess twisting and shorter than 150-bp, the length regime where the free energy of loop formation is dominated by bending energy.

We first tried the loop capture geometry used in DNA cyclization, which we term as the ``hairband loop" (Figure \ref{fig2}(a)). In this geometry, the complementary overhangs protrude from different strands so that the sticky ends can anneal in trans and stack upon each other. In a previous study, we showed that this end stacking, or equivalently nick closing, substantially increases the hairband loop stability \cite{Jeo}. Using the single-molecule FRET assay, we measured the hairband loop capture times with and without base pair mismatch in the center. As shown in Figure \ref{fig2}(b), hairband loop capture took less time in the presence of the mismatch as expected. The loop capture time further decreased with increasing mismatch size (circles, Figure \ref{fig2}(b)). The base pair mismatch in the center position led to the largest decrease in the loop capture time, and the decrease dropped as the mismatch was placed further from the center (triangles, Figure \ref{fig2}(b)). %The measured capture time also changed dramatically with the size and position of the mismatch (Figure \ref{fig2}(b)). 
These observations confirm previous findings that mismatched base pairs reduce the energy barrier for loop formation by increasing DNA bendability \cite{Sharma2013, Kahn1994, Schallhorn2005, Yuan2006}, and this barrier reduction is most effective when the mismatch is in the center \cite{Ranjith2005}.

Next, we measured the hairband loop release times or loop lifetimes ($\tau_{\textrm{loop}}$) with and without the mismatch in the center. Since a mismatch could relieve the bending stress of the hairband loop, we thought that the loop lifetime would become longer. To our surprise, we observed the exact opposite effect where the central mismatch decreased the hairband loop lifetime (Figure \ref{fig2}(c)). Increasing the size of the mismatch from 1 bp to 3 bp led to a further decrease in the lifetime. This effect seemed to plateau past the mismatch size of 3 bp (Figure \ref{fig2}(c)). This result suggests that the mismatch-containing hairband loop is more kinetically unstable than the mismatch-free loop, which seems paradoxical through the lens of the one-dimensional model presented in Figure \ref{fig1}(a).

We thus considered the possibility that the transition state depends on other reaction coordinates besides the end-to-end distance, such as the closing angles at the loop junction. Since base stacking at the nick(s) in the hairband loop is a key determinant of decyclization kinetics \cite{Jeo}, we asked whether the central mismatch could destabilize the hairband loop by allosterically inducing nick opening. To investigate such allosteric coupling, we calculated the curvature profile of a kinkable semiflexible loop \cite{Vologodskii2013} containing a defect with zero rigidity from a Monte Carlo simulation (see Supplemental Material for details). As shown in Figure \ref{fig2}(d), a kink with a sharp bending angle appeared most frequently at the furthest end of the loop from the defect. We also calculated the minimum energy conformation of a semiflexible loop while varying the rigidity of the defect and found that the bending angles of furthest points were highly correlated (Figure \ref{fig2}(e)). This loop-mediated correlation of sharp bending angles between most distant sites is termed cooperative kinking \cite{Lionberger2011}, and has been observed in torsionally strained DNA minicircles by cryo-electron microscopy and molecular dynamics simulations \cite{Lionberger2011,irobalieva2015structural,sutthibutpong2016long}. 

\begin{figure}[b]
\begin{center}
\includegraphics[scale = 1]{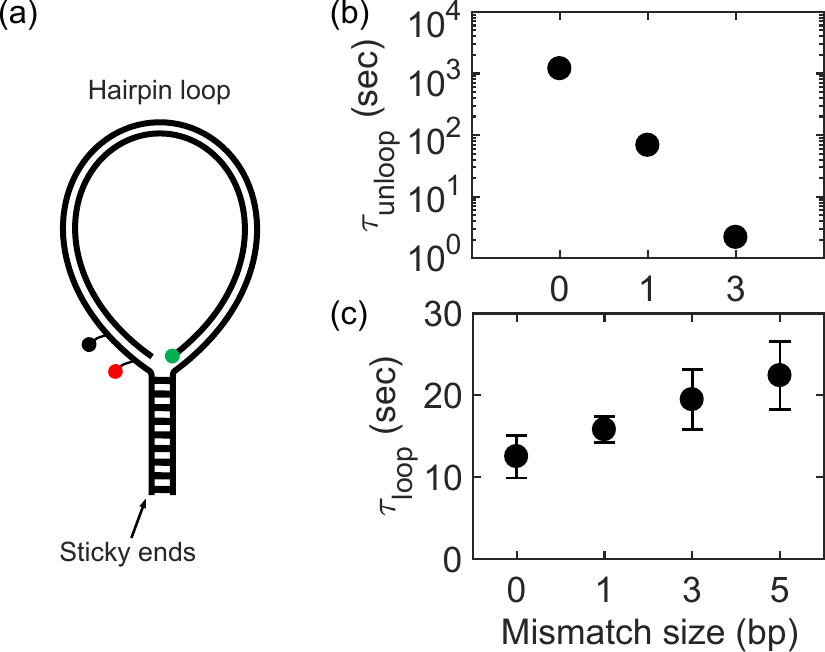}
\end{center}
\caption{\textbf{(a)} Schematic of a hairpin loop. The schematic shows the FRET pair (green and red circles), the biotin linker (black circle), and base-paired overhangs. In this geometry, the overhangs on the same strand form a duplex like a zipper. \textbf{(b)} Loop capture time of the hairpin (105 bp) molecules as a function of the central mismatch size. Error bars are omitted due to their small sizes. \textbf{(c)} Hairpin loop lifetime as a function of the central mismatch size. Error bars represent the standard errors of the mean.}
\label{fig3}
\end{figure}

\begin{figure}[!t]
\begin{center}
\includegraphics[scale = 1]{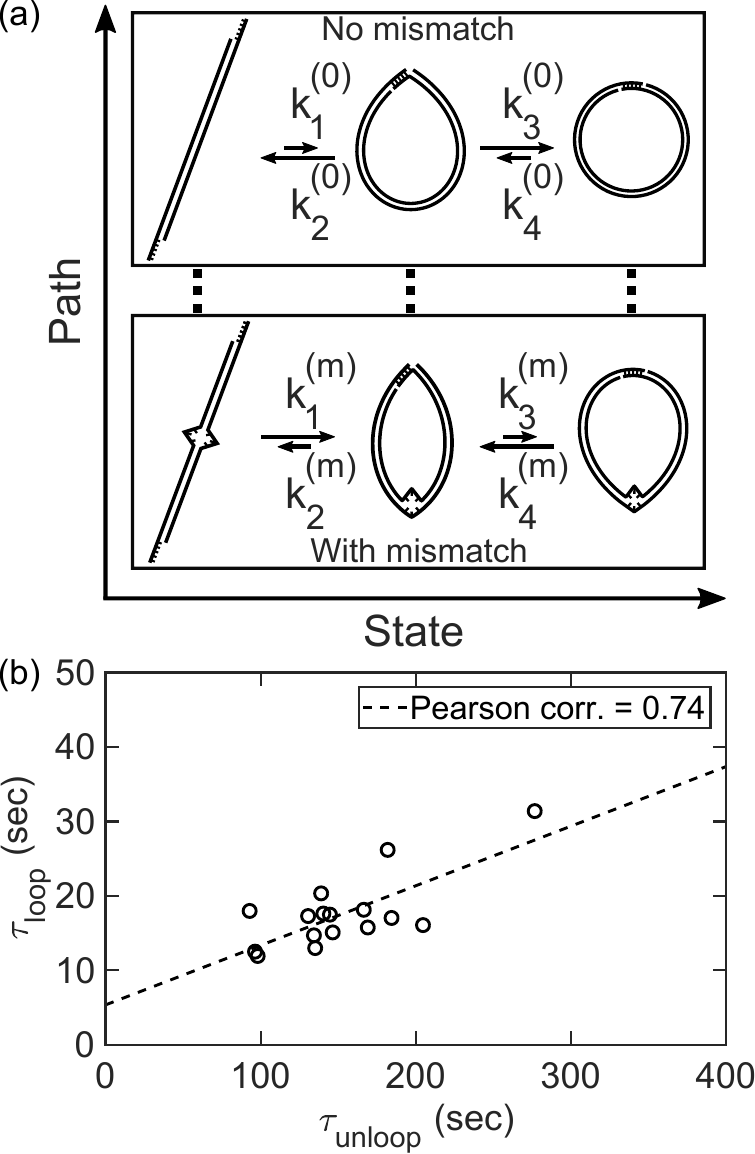}
\end{center}
\caption{\textbf{(a)} The three-state model for hairband loop closure and release. The three states from left to right are unlooped, unstacked, and stacked states. The looped state is a mixed state between the unstacked and stacked states. Therefore, the apparent loop capture rate ($k_\textrm{loop}$) is equal to $k_1$, but the apparent loop release rate ($k_\textrm{unloop}$) depends on $k_2$, $k_3$, and $k_4$. For the hairpin loop, $k_3=0$, and therefore, $k_\textrm{unloop}$ is equal to $k_2$. Two representative paths for central mismatch size $0$ and $m$ are highlighted with arc-like (top) and tweezers-like (bottom) motions, respectively. The vertical dotted lines imply the continuum of paths running parallel to the two extreme ones shown. \textbf{(b)} Correlation between loop capture and release times of 16 unrelated hairband DNA molecules of the same size (94bp). The loop capture and release times were measured in equilibrium (i.e. no buffer-exchange) at slightly elevated temperature of 34 $^\circ$C with [NaCl] = 700mM.}
\label{fig4}
\end{figure}
We hypothesized that the enhanced flexibility of the central mismatch destabilizes the hairband loop preventing nicks(s) on the opposite side from closing. This hypothesis provides a few testable predictions. First, if the mismatch were displaced from the midpoint of the DNA, the degree of destabilization would be dampened. In agreement with this prediction, we observed a longer loop lifetime when the mismatch was placed at a quarterpoint instead of the center (Figure \ref{fig2}(f)). Second, the cooperative kinking hypothesis requires nicks that can buckle under the bending stress, and therefore the mismatch-induced destabilization would be eliminated in a loop capture geometry free of end-stacking. We thus tested a different loop geometry referred to as the ``hairpin loop", where the complementary overhangs protrude from the same strand (Figure \ref{fig3}(a)). In this geometry, the sticky ends anneal in cis and cannot stack upon each other. Using these new DNA constructs with a central mismatch of various sizes, we repeated loop capture and release experiments. Similar to hairband loop capture, the hairpin capture time decreased with the size of base pair mismatch (Figure \ref{fig3}(b)). However, in sharp contrast to the hairband loop, the hairpin loop lifetime increased with mismatch size (Figure \ref{fig3}(c)). The effect of the base pair mismatch on the hairpin loop stability is therefore consistent with the prediction of the one-dimensional model. Overall, the lifetimes of hairpin loops were shorter than those of hairband loops, which is consistent with easier rupture of DNA duplex in an unzipping geometry than in a shearing geometry \cite{Mosayebi2015,Zhang2016,Tee2018}. These results lend strong support to the idea that cooperative kinking governs the kinetic stability of a mismatch-containing hairband loop.

The mismatch-dependence of the hairband loop release kinetics reveals the limitations of the one-dimensional two-state model (Figure \ref{fig1}(a)) and invites us to consider additional states and alternative reaction paths along another dimension. Here, we present two different paths ($k^{(0)}$ and $k^{(m)}$) that are likely to be the dominant ones for mismatch-free and mismatch-containing DNA (Figure \ref{fig4}(a)). Each path goes through three different states: unlooped, unstacked, and stacked. The loop capture rate is much greater in the presence of a central mismatch due to its enhanced flexibility ($k^{(m)}_1 \gg k^{(0)}_1$). The reverse rate is expected to be slower with the mismatch ($k^{(\textrm{m})}_2 < k^{(\textrm{0})}_2$) because of the weaker loop tension. Mismatch-free DNA undergoes small bending fluctuations uniformly throughout its contour, and therefore, follows an arc-like trajectory toward the looped state where end-stacking (nick closing) and end-unstacking (nick opening) transitions may occur. In comparison, DNA with a mismatch in the center can be sharply bent at a much lower energy cost, and therefore, the most dominant path toward the looped state will resemble a tweezers-like motion. As a result of this motion, the sticky ends anneal at a sharp angle, and the hairband loop with the mismatch faces a higher energy barrier for end-stacking (nick closing) than without ($k^{(m)}_3 \ll k^{(0)}_3$). The mismatch not only suppresses end-stacking, but also promotes end-unstacking (nick opening) through cooperative kinking, which implies $k^{(m)}_\textrm{4} \gg k^{(0)}_\textrm{4}$. Hence, the apparent release rate of the hairband loop ($k_\textrm{unloop}$) becomes faster with the mismatch than without because the looped state with the mismatch is heavily biased towards the unstacked state. In comparison, for the hairpin loop that cannot proceed to the stacked state, the three-state model is reduced to the two-state model, and the loop release rate is slower with the mismatch ($k^{(\textrm{m})}_2 < k^{(\textrm{0})}_2$). 

%Hence, the apparent loop release rate, $k_\textrm{unloop} = k_\textrm{2}\cdot \frac{k_\textrm{4}}{k_\textrm{3}+k_\textrm{4}}$, reduces to $k^{(m)}_\textrm{2}$ for large $m$. Although unbending may be slower with mismatch ($k^{(m)}_\textrm{2} < k^{(0)}_\textrm{2}$), the two-path model suggests that the loop release can still be faster because of the central mismatch blocking transition to the looped state with closed nicks, which explains the observed positive correlation between the loop capture and release times as a function of mismatch size.
The two paths boxed in Figure 4(a) represent the two most extreme paths in terms of kinetics, the top path for the slowest hairband loop capture and release, and the bottom for the fastest. In reality, there exists a continuum of paths going through the three states with intermediate rates, and the flexibility profile of DNA determines the relative weights at which individual paths are taken. Therefore, any changes to the flexibility profile of DNA would lead to correlated changes in the hairband loop capture and release rates. To test this idea, we measured hairband loop capture and release times of 16 unrelated sequences, all of the same length. Although limited in sample size, we observed a significant degree of correlation between the two times (Pearson correlation = 0.74, Figure \ref{fig4}(b)). This result suggests that cooperative kinking is a general mechanism that governs the kinetics of hairband loop capture and release. 

In conclusion, we demonstrate that base pair mismatch can constrain the geometry and interactions for DNA loop capture through cooperative kinking, and the close coupling between hairband loop geometry and end-stacking can give rise to correlated changes between loop capture and release times (``easy come, easy go''). We propose a three-state model that correctly describes the effect of mismatched base pairs on the apparent kinetics of loop capture and release. We expect the effect of mismatched base pairs on protein-mediated DNA loops to be more complex because of the diversity in loop capture geometry \cite{haeusler2012fret}. Beyond passively captured DNA loops, it would be interesting to investigate whether base pair mismatches can also influence the kinetics of DNA loop extrusion \cite{Ganji2018, Marko325373} through cooperative kinking.

\bibliography{references}
\end{document}

%\preprint{APS/123-QED}

\title{Supplemental Material to ``Base-pair mismatch can destabilize small DNA loops through cooperative kinking''}% Force line breaks with \\
%new title needed

%Cooperative kinking in DNA cyclization
%Easy come, easy go: loops 

\author{Jiyoun Jeong}
\author{Harold D. Kim}%
 \email{Corresponding author.\\Email: harold.kim@physics.gatech.edu}
\affiliation{%
 School of Physics, Georgia Institute of Technology, 837 State Street, Atlanta, GA 30332-0430, USA
}%

\date{\today}% It is always \today, today,

\pacs{Valid PACS appear here}% PACS, the Physics and Astronomy
\maketitle

%\tableofcontents
\section{Materials and methods}
\subsection{Preparation of DNA molecules}
A 105-bp-long DNA molecule was extracted from yeast genomic DNA by polymerase chain reaction (PCR) to serve as a control DNA template without any structural defect. To probe the effect of a permanent defect, we planned to introduce a DNA mismatch to the control molecule by mixing it with its mutant followed by a strand-exchange reaction. To do so, we additionally prepared a set of mutated DNA molecules that differ from the control only in a certain location in which we put a mutation of size equal to 1bp, 3bp, or 5bp. To make such molecules, first, the mutated templates of the control DNA were synthesized from Eurofins Genomics (EXTREMer oligos) and duplexed via PCR. Each of the duplexed products was then incorporated into a pJET1.2$\backslash$blunt vector (ThermoFisher) and cloned into DH5\textrm{$\alpha$} \textit{Escherichia coli} cells. Finally, the cloned fragments of DNA were extracted via colony PCR from the cells and were sequenced to ensure the correct mutation was made at the desired location.

To modify these molecules to carry a FRET pair (i.e. Cy3 and Cy5), biotin, and single-stranded sticky ends, we followed our standard preparation protocol \cite{Le2014jove}, which involves a series of PCR and strand exchange reactions that can be found in elsewhere. For introducing a DNA mismatch in the final construct, we mixed the Cy3-labled control molecule with one of the Cy5-labeled mutated molecules with a ratio of 4:1 in the strand-exchange reaction. 

The final DNA construct generated by this protocol carries a 5$^\prime$ protruding sticky end on each end and makes a hairband loop upon end-annealing as shown in Figure 2(a) of the main text. We also made hairpin loops by having sticky ends on the same DNA strand (Figure 3(a) of the main text). A complete list of all DNA sequences can be found in Tables \ref{supp-tab1} and \ref{supp-tab2} below.

\subsection{single-molecule FRET looing and unlooping assay}
We followed our previous single-molecule FRET assay that employs the sudden salt-exchange protocol \cite{Le2014,Jeo}. For cyclization, DNA molecules were deposited on a passivated surface of a flow-cell and were incubated at a low salt (10 mM [NaCl) imaging buffer containing the PCD-PCA oxygen scavenging system \cite{Aitken2008} for 10 minutes. We then injected a high salt (1 M [NaCl]) imaging buffer into the flow-cell to promote sticky ends to capture the loop configuration. Decyclization measurements were done similarly, except that the NaCl concentration was changed from 2 M to 75 mM. The immobilized molecules were excited by a 532-nm laser continuously through an objective-type TIR microscope from the beginning of the buffer exchange. The time trajectories of FRET signals (Figure 1(b) of the main text) from the molecules were recorded by an EMCCD camera (DU-897ECS0-\# BV, Andor) at a rate of 100 ms per frame for the mismatch-free molecules and 50 ms per frame for the molecules with a mismatch.

\subsection{Minicircle simulations}
The Monte Carlo simulation of a minicircle was implemented as previously described \cite{zheng2009theoretical,Le2014}. A set of 105 connected nodes was used to create a coarse-grained representation of a DNA minicircle of 105 bp. The bending energy at each node was described by the kinkable worm-like chain model \cite{zheng2009theoretical} with the parameters of b = 0.3 and h = 12 following the same notation used in Ref. \cite{Vologodskii2013a}. We performed the simulation with and without a flexible defect of zero bending energy placed at a fixed location. For the case of no flexible spot, we first initialized the simulation without allowing the kink formation. Once the kink-free simulation was equilibrated, we allowed spontaneous kinks to appear. To construct the probability density of kink positions, we ran the simulation and stop at the first appearance of a kink. We then recorded the position of this kink and equilibrated back to the kink-free state. This procedure was repeated until we collected a distribution of 1000 kink positions. The same procedure was repeated in the presence of the hyperflexible spot to predict the effect of a flexible spot on the probability distribution of kink.

\clearpage
\onecolumngrid
\begin{longtable*}{| P{0.14\textwidth}| p{0.75\textwidth}| }

\caption{DNA sequences of hairband molecules.}\label{supp-tab1} \\
\hline \hline
%\multicolumn{2}{|c|}{\begin{tabular}[c]{@{}l@{}}\textbf{Hairpin molecules}\end{tabular}}
%\\ \hline
\endfirsthead

\hline \multicolumn{2}{|r|}{{Continued on next page}} \\ \hline
\endfoot

\endlastfoot

No mismatch & 5$^\prime-$\underline{TGAATTTAC}\seqsplit{G}\textcolor{red}{\textbf{T}}\seqsplit{GCCAGCAACAGA[T]AGCCGCGATCGCCATGGCAACGAGGTCGCACACGCCCCACACCCAGACCTCCCTGCGAGCGGGCATGGGTACAATCATTCGAGCTCGTTGTAG}-3$^\prime$ \newline
3$^\prime-$\seqsplit{CACGGTCGTTGTCTATCGGCGCTAGCGGTACCGTTGCTCCAGCGTGTGCGGGGTGTGGGTCTGGAGGGACGCTCGCCCGTACCCATGTTAGTAAGCTCGAGCAACA}\textcolor{green}{\textbf{T}}C\underline{ACTTAAATG}-5$^\prime$
\\ \hline
1bp-mismatch (central) & 5$^\prime-$\underline{TGAATTTAC}\seqsplit{G}\textcolor{red}{\textbf{T}}\seqsplit{GCCAGCAACAGA[T]AGCCGCGATCGCCATGGCAACGAGGTCGCACACGCCCCAGACCCAGACCTCCCTGCGAGCGGGCATGGGTACAATCATTCGAGCTCGTTGTAG}-3$^\prime$ \newline
3$^\prime-$\seqsplit{CACGGTCGTTGTCTATCGGCGCTAGCGGTACCGTTGCTCCAGCGTGTGCGGGGTCTGGGTCTGGAGGGACGCTCGCCCGTACCCATGTTAGTAAGCTCGAGCAACA}\textcolor{green}{\textbf{T}}C\underline{ACTTAAATG}-5$^\prime$
\\ \hline
3bp-mismatch (central) & 5$^\prime-$\underline{TGAATTTAC}\seqsplit{G}\textcolor{red}{\textbf{T}}\seqsplit{GCCAGCAACAGA[T]AGCCGCGATCGCCATGGCAACGAGGTCGCACACGCCCCGGGCCCAGACCTCCCTGCGAGCGGGCATGGGTACAATCATTCGAGCTCGTTGTAG}-3$^\prime$ \newline
3$^\prime-$\seqsplit{CACGGTCGTTGTCTATCGGCGCTAGCGGTACCGTTGCTCCAGCGTGTGCGGGGCCCGGGTCTGGAGGGACGCTCGCCCGTACCCATGTTAGTAAGCTCGAGCAACA}\textcolor{green}{\textbf{T}}C\underline{ACTTAAATG}-5$^\prime$
\\ \hline
5bp-mismatch (central) & 5$^\prime-$\underline{TGAATTTAC}\seqsplit{G}\textcolor{red}{\textbf{T}}\seqsplit{GCCAGCAACAGA[T]AGCCGCGATCGCCATGGCAACGAGGTCGCACACGCCCGCGCGCCAGACCTCCCTGCGAGCGGGCATGGGTACAATCATTCGAGCTCGTTGTAG}-3$^\prime$ \newline
3$^\prime-$\seqsplit{CACGGTCGTTGTCTATCGGCGCTAGCGGTACCGTTGCTCCAGCGTGTGCGGGCGCGCGGTCTGGAGGGACGCTCGCCCGTACCCATGTTAGTAAGCTCGAGCAACA}\textcolor{green}{\textbf{T}}C\underline{ACTTAAATG}-5$^\prime$
\\ \hline
3bp-mismatch (10 bp off-center)& 5$^\prime-$\underline{TGAATTTAC}\seqsplit{G}\textcolor{red}{\textbf{T}}\seqsplit{GCCAGCAACAGA[T]AGCCGCGATCGCCATGGCAACGAGGTCGTGGACGCCCCACACCCAGACCTCCCTGCGAGCGGGCATGGGTACAATCATTCGAGCTCGTTGTAG}-3$^\prime$ \newline
3$^\prime-$\seqsplit{CACGGTCGTTGTCTATCGGCGCTAGCGGTACCGTTGCTCCAGCACCTGCGGGGTGTGGGTCTGGAGGGACGCTCGCCCGTACCCATGTTAGTAAGCTCGAGCAACA}\textcolor{green}{\textbf{T}}C\underline{ACTTAAATG}-5$^\prime$
\\ \hline
3bp-mismatch (20 bp off-center) & 5$^\prime-$\underline{TGAATTTAC}\seqsplit{G}\textcolor{red}{\textbf{T}}\seqsplit{GCCAGCAACAGA[T]AGCCGCGATCGCCATGGCGGTGAGGTCGCACACGCCCCACACCCAGACCTCCCTGCGAGCGGGCATGGGTACAATCATTCGAGCTCGTTGTAG}-3$^\prime$ \newline
3$^\prime-$\seqsplit{CACGGTCGTTGTCTATCGGCGCTAGCGGTACCGCCACTCCAGCGTGTGCGGGGTGTGGGTCTGGAGGGACGCTCGCCCGTACCCATGTTAGTAAGCTCGAGCAACA}\textcolor{green}{\textbf{T}}C\underline{ACTTAAATG}-5$^\prime$
\\ \hline
\end{longtable*}

\begin{longtable*}{| P{0.14\textwidth}| p{0.75\textwidth}| }
\caption{DNA sequences of hairpin molecules.}\label{supp-tab2} \\
\hline \hline
%\multicolumn{2}{|c|}{\begin{tabular}[c]{@{}l@{}}\textbf{Hairpin molecules}\end{tabular}}
%\\ \hline
\endfirsthead

\hline \multicolumn{2}{|r|}{{Continued on next page}} \\ \hline
\endfoot
%\endhead
\endlastfoot
%\hline \multicolumn{2}{|r|}{{Continued on next page}} 
%\multicolumn{2}{c}{{\bfseries \tablename\ \thetable{} -- continued from previous page}} \\
%\hline 
%\multicolumn{2}{|c|}{\begin{tabular}[c]{@{}l@{}}\textbf{DNA set 1 (5\textrm{$^\prime$} to 3\textrm{$^\prime$})}\end{tabular}} \\ 
%\hline 
%\endhead

No mismatch & 5$^\prime-$\underline{TGAATTTACG}(CT)G\textcolor{red}{\textbf{T}}\seqsplit{GCCAGCAACAGA[T]AGCCACATCGCCATGGCAACGAGGTCGCACACGCCCCACACCCAGACCTCCCTGCGAGCGGGCATGGGTTGCATGTCAGCTATGGATCCATTCGTAAATTCA}-3$^\prime$ \newline
3$^\prime-$\seqsplit{CACGGTCGTTGTCTATCGGTGTAGCGGTACCGTTGCTCCAGCGTGTGCGGGGTGTGGGTCTGGAGGGACGCTCGCCCGTACCCAACGTACAGT}(CG)\underline{ATACCTAGGT}-5$^\prime$[\textcolor{green}{Cy3}]
\\ \hline
1bp-mismatch (central) & 5$^\prime-$\underline{TGAATTTACG}(CT)G\textcolor{red}{\textbf{T}}\seqsplit{GCCAGCAACAGA[T]AGCCACATCGCCATGGCAACGAGGTCGCACACGCCCCAGACCCAGACCTCCCTGCGAGCGGGCATGGGTTGCATGTCAGCTATGGATCCATTCGTAAATTCA}-3$^\prime$ \newline
3$^\prime-$\seqsplit{CACGGTCGTTGTCTATCGGTGTAGCGGTACCGTTGCTCCAGCGTGTGCGGGGTCTGGGTCTGGAGGGACGCTCGCCCGTACCCAACGTACAGT}(CG)\underline{ATACCTAGGT}-5$^\prime$[\textcolor{green}{Cy3}]
\\ \hline
3bp-mismatch (central) & 5$^\prime-$\underline{TGAATTTACG}(CT)G\textcolor{red}{\textbf{T}}\seqsplit{GCCAGCAACAGA[T]AGCCACATCGCCATGGCAACGAGGTCGCACACGCCCCGGGCCCAGACCTCCCTGCGAGCGGGCATGGGTTGCATGTCAGCTATGGATCCATTCGTAAATTCA}-3$^\prime$ \newline
3$^\prime-$\seqsplit{CACGGTCGTTGTCTATCGGTGTAGCGGTACCGTTGCTCCAGCGTGTGCGGGGCCCGGGTCTGGAGGGACGCTCGCCCGTACCCAACGTACAGT}(CG)\underline{ATACCTAGGT}-5$^\prime$[\textcolor{green}{Cy3}]
\\ \hline
5bp-mismatch (central) & 5$^\prime-$\underline{TGAATTTACG}(CT)G\textcolor{red}{\textbf{T}}\seqsplit{GCCAGCAACAGA[T]AGCCACATCGCCATGGCAACGAGGTCGCACACGCCCGCGCGCCAGACCTCCCTGCGAGCGGGCATGGGTTGCATGTCAGCTATGGATCCATTCGTAAATTCA}-3$^\prime$ \newline
3$^\prime-$\seqsplit{CACGGTCGTTGTCTATCGGTGTAGCGGTACCGTTGCTCCAGCGTGTGCGGGCGCGCGGTCTGGAGGGACGCTCGCCCGTACCCAACGTACAGT}(CG)\underline{ATACCTAGGT}-5$^\prime$[\textcolor{green}{Cy3}]
\\ \hline

\end{longtable*}
\begin{minipage}{0.95\textwidth}
Both top (5$^\prime$ to 3$^\prime$) and bottom (3$^\prime$ to 5$^\prime$) sequences are shown. The underlined sequences represent sticky ends. A Cy5 fluorophore is internally attached at the thymine base colored in red. A Cy3 fluorophore is either at the green thymine base or the 5$^\prime$ end of the bottom strand. A biotin molecule is linked to the thymine base shown as [T]. Hairpin molecules includes a 2-nt gap (indicated by sequences in parentheses) near each end of the top strand before sticky ends.
\end{minipage}

\bibliography{supplementary-refs}